\documentclass[12pt,preprint]{aastex}

\slugcomment{Accepted for publication in ApJ}

\shorttitle{Molecular shells in IRC+10216}
\shortauthors{Dinh-V-Trung}

\begin{document}

\title{Molecular shells in IRC+10216: Evidence for non-isotropic and epsiodic mass loss enhancement}

\author{Dinh-V-Trung\footnote{on leave from Center for Quantum Electronics, Institute of Physics 
P.O Box 423, Bo Ho 10000, Hanoi, Vietnam}}
\affil{Institute of Astronomy and Astrophysics, Academia Sinica\\ 
P.O Box 23-141, Taipei 106, Taiwan.}
\email{trung@asiaa.sinica.edu.tw}
\author{Jeremy Lim}
\affil{Institute of Astronomy and Astrophysics, Academia Sinica\\ 
P.O Box 23-141, Taipei 106, Taiwan.}
\email{jlim@asiaa.sinica.edu.tw}

\begin{abstract}
We report high angular-resolution VLA
observations of cyanopolyyne molecules HC$_3$N and HC$_5$N from the carbon rich 
circumstellar envelope of IRC+10216. The observed low-lying rotational transitions trace 
a much more extended emitting region than seen in previous observations at higher frequency transitions. We resolve the 
hollow quasi-spherical distribution
of the molecular emissions into a number of clumpy shells. These molecular shells coincide spatially
with dust arcs seen in deep optical images of the IRC+10216 envelope, allowing us to study for the
first time the kinematics of these features. We find that the molecular
and dust shells represent the same density enhancements in the envelope separated in time by $\sim$120 to $\sim$360 yrs. 
From the angular size and velocity spread of the shells, we estimate that each shell typically covers
about 10\% of the stellar surface at the time of ejection.
The distribution of the shells seems to be random in space. 
The good spatial correspondance between HC$_3$N and HC$_5$N emissions is in qualitative agreement with a recent
chemical model that takes into account the presence of density-enhanced shells. The broad spatial distribution
of the cyanopolyyne molecules, however, would necessitate further study on their formation.
\end{abstract}


\section{Introduction}
Near the end of their lifes, 
intermediate-mass stars (1 M$_\odot$ $\leq$ M$_*$ $\leq$ 8 M$_\odot$) evolve through the asymptotic giant branch (AGB),
which is characterised by copious mass loss from the stars in the form of a slow and dusty wind driven by stellar radiation pressure
on dust particles. It has been commonly assumed that
the mass loss from AGB stars is more or less isotropic and steady over time. Recent optical imaging
(Sahai et al. 1998, Hrivnak et al. 2001, Schmidt et al. 2002), however, reveals the presence of concentric 
arcs and/or rings of dust in the circumstellar envelope around a number of AGB and also post-AGB stars. 
These studies suggest that the mass loss from AGB stars, although approximately 
isotropic, is modulated on a timescale of a few hundreds years. 
This timescale is much longer than typical stellar pulsation periods but much shorter than the
interval between helium flashes. Several possibilities have been advanced to 
explain these arcs such as the modulation of mass loss by dust-gas coupling effect (Simis et al. 2001) or magnetic cycle (Soker 2000).
A better understanding of the physical properties of these features is needed before we can draw firm conclusions
on their formation mechanisms.

IRC+10216 (CW Leo) is a nearby AGB star with a high mass loss rate, estimated at 3x10$^{-5}$ M$_\odot$ yr$^{-1}$. The distance
to this star is estimated to be in the range 110 - 150 pc (Groenewegen 1998, Lucas \& Gu\'{e}lin 1999). We adopt in this paper a 
distance of 120 pc, which is within the range of distances to give good agreement between model results and the observations of CO rotational lines 
(Groenewegen 1998). 
The combination of proximity and the high mass loss rate makes IRC+10216 one of the strongest sources molecular line emissions
in the sky.
Deep optical images taken by Mauron \& Huggins (1999) and more recently by Le\~{a}o et al. (2006) show that 
the dusty envelope is not smooth but consists of
a series of arcs or incomplete shells. The average angular separation between the dust arcs 
suggests a timescale for the change in mass loss rate of order 200 to 800 yrs.
 
We emphasize here that the lack of kinematic information on the dust arcs precludes any firm conclusion on
the real 3-dimension structure of the arcs or shells.
From large scale mapping at a relatively low angular resolution ($\sim$ 12 arcsec) of the CO J=1$-$0 emission from the envelope of
IRC+10216, Fong et al. (2003) discovered a series of large molecular arcs or shells
(at radii of $\sim$100 arcsecs) in the outer envelope. They attribute these arcs to
reverberations in the envelope generated 
by a previous helium flash. The timescale inferred from the spacing between these arcs is about 1000 - 2000 years. In
addition, they suggest that the dust arcs seen in optical images are actually projections on the
plane of the sky of these molecular arcs observed in CO J=1$-$0, even though the dust arcs 
are found much closer ($\sim$20 to 60 arcsec) to the central star.

The molecular envelope of IRC+10216 has been imaged extensively at high angular resolution (as high as 3 arcsec)
using interferometers operating at 3mm wavelengths
(see Lucas \& Gu\'{e}lin 1999 for a review). The distribution of different molecules show a striking
dichotomy: for the molecules originating from the inner region of the envelope, i.e SiO, CS and SiS show a compact centrally peaked
distribution, whereas more complex molecules such as carbon chain molecules show hollow quasi-spherical
distribution ranging from 15 to 20 arcsecs in radius 
(Bieging \& Nguyen-Q-Rieu 1988, Bieging \& Tafalla 1993, and Lucas \& Gu\'{e}lin 1999). 
This morphological differences 
have been attributed to the active photochemistry and molecule-radical reactions in the envelope of IRC+10216 (Cherchneff et al. 1993).
The presence of many different molecules distributed over a large spatial extent of the envelope has been used to probe the 
structure and dynamics of the envelope, but until now the observations have not had sufficient sensitivity and angular resolution
to trace molecular emissions associated with the individual dust arcs or shells. 

In this paper, we present observations of the cyanopolyyne molecules HC$_3$N and HC$_5$N from IRC+10216 
obtained with the Very Large Array (VLA\footnote{The VLA is a facility of the
National Radio Astronomy Observatory, which is operated by Associated Universities, Inc., under
contract with the National Science Foundation}).
The VLA allows us to image the molecular emissions from the envelope of IRC+10216 at unprecedented 
angular resolution of $\sim$1.5 arcsec. Our observations reveal the presence of mutiple incomplete molecular shells 
and larger spatial extent of the cyanopolyyne molecules than previously seen. 
\section{Observation}
We observed IRC+10216 ($\alpha_{\rm J2000}$=09:47:57.38, $\delta_{\rm J2000}$=13:16:43.7) using the VLA in its
most compact configuration (D-array) on 2004, April 12 and 13 under good weather condition. 
The phase center chosen in our observations coincides with the position of the central compact 
continuum source at 3mm in IRC+10216 (Gu\'{e}lin et al. 1993). This position is also consistent with the position of the
central continuum source at centimeter wavelengths in IRC+10216 as measured by Drake et al. (1991) and Menten et al. (2006).
The rest frequency of the HC$_3$N J=$5-$4 line (45.490316 GHz) is taken from Lovas/NIST database (Lovas 2004). The
rest frequencies of HC$_5$N J=$9-$8 (23.963897 GHz) and J=$16-$15 (42.602171 GHz) rotational transitions are taken
from JPL catalogue (Pickett et al. 1998). The VLA correlator 
was configured to cover a bandwidth
of 6.25 MHz with a frequency resolution of 97.656 kHz over 64 channels in the two-IF mode for the HC$_3$N J=$5-$4 line. 
We used the 4-IF mode to observe simultaneously the HC$_5$N J=$16-$15 line and the nearby HC$_7$N J=$38-$37 line 
with a frequency resolution of 195.313 KHz over 32 channels. We found the
HC$_7$N line too weak to image, and will not be discussed any further in this paper. For the HC$_5$N J=9$-$8 line 
at 1.3cm, we used the correlator in the 4-IF mode covering a bandwidth of 3.125 MHz over 64 channels.
 
Nearby quasar 0854+201 was observed 
every 15 to 20 minutes to correct for time-dependent variations in the antennae gains. The strong quasar 
1229+020 was used to correct
for the shape of the antennae passband. The absolute flux scale was determined using the standard quasar 1331+305.
At 7mm ($\sim$43 GHz), the field of view of the antennas is approximately 60 arcsecs, comparable with the size of
the emitting region. To cover the whole emitting region, we therefore used a 7-pointing (hexagonal pattern) 
mosaic with 20 arcsecs separation between pointings. At 1.3cm, the field of view of approximately 120 arcsec is
sufficient to cover the whole emitting region. We therefore used only a single pointing for observations at 1.3cm.
We edited and calibrated the visibilities using the AIPS package. We then exported the calibrated
visibilities into the MIRIAD package for imaging and deconvolution. To form the
mosaiced images at 7mm, we assumed the primary beam of VLA antennas to be a two-dimensional Gaussian with a FWHM of 60 arcsecs. 
We applied an inverse Fourier transform of the calibrated visibilities to form the mosaiced images, and then deconvolved 
the point spread function of the interferometer using the 
Steer-Dewdney-Ito clean algorithm (Steer et al. 1984). For the HC$_5$N J=$9-$8 line 
at 1.3cm, we deconvolved the single-field image using the clean algorithm of Hogbom, Clark and Steer as implemented
in the task CLEAN of the MIRIAD package. A summary of our observations is 
shown in Table. 1.
\section{Results}
We show in Figure 1. the spatial distribution of the emissions from the HC$_3$N J=$5-$4, HC$_5$N J=16$-$15 and HC$_5$N J=9$-$8 lines
in three velocity channels around the systemic
velocity of V$_{\rm LRS}$=$-$26 kms$^{-1}$ together with the optical image of Le\~{a}o et al. (2006), which is color-coded
to highlight the presence of numerous dust arcs in the envelope of IRC+10216. 
Because we did not perform continuum subtraction for our data,
continuum emission, which has a flux density of about 3 mJy at 1.3cm, appears at the phase center position in 
the HC$_5$N J=9$-$8 channel maps at the level of $\sim$3$\sigma$. The central continuum source is not detectable in the
HC$_3$N J=5$-$4 and HC$_5$N J=16$-$15 channel maps observed at 7mm. In Figure 2 we show the total flux profiles
of the observed transitions. The integration is done within the emitting region of cyanopolyynes, taking into
account all pixels with flux above 2$\sigma$ level. The HC$_3$N J=5$-$4 and HC$_5$N J=16$-$15 transitions were observed  previously
with the Nobeyama 45m telescope (Kawaguchi et al. 1995), which has a FWHM beam of 40 arcsec. Using the main beam efficiency provided
by Kawaguchi et al. (1995) and a conversion factor of $\sim$3 Jy/K, the single dish fluxes of HC$_3$N J=5$-$4 and HC$_5$N J=16$-$15
transitions at the systemic velocity are 6 Jy and 1.5 Jy, respectively. Because most of the emitting region 
of cyanopolyynes is within the beam of the 45m telescope, we estimate that in our VLA observations 
we recover most of the flux of HC$_3$N J=5$-$4 transition and about 70\% of the flux of HC$_5$N J=16$-$15 transition.
The HC$_5$N J=9$-$8 transition was observed by Bell et al. (1992) using the 140 foot telescope. The FWHM telescope beam
at the frequency of this transition is 1$'$.4. Using their quoted value of the beam efficiency and a conversion factor of 
$\sim$3 Jy/K, the flux of the HC$_5$N J=9$-$8 transition is about 400 mJy. By comparison with the total flux profile of this
transition shown in Figure 2, we estimate that in our VLA observations we recover more than 70\% of the single dish flux.
We note, however, that lacking a fully sampled single dish 
map of the cyanopolyyne emission prevents us from estimating precisely the amount 
of flux recovered.

The emissions of both molecules have a clumpy and hollow quasi-spherical distribution, which is
also seen previously in higher frequency transitions (Bieging \& Nguyen-Q-Rieu 1989, Bieging \& Tafalla 1993 and
Lucas \& Gu\'{e}lin 1999). Because HC$_3$N is thought to form mainly from the reaction between radical CN, which
is a photodissociation product of parent molecule HCN, and acetylene molecule:
$
CN + C_2H_2 \longrightarrow HC_3N + H
$, while bigger cyanopolyynes are built up step by step from smaller chains,
the chemical models for carbon-rich circumstellar envelopes such
as IRC+10216 (Cherchneff et al. 1993, Millar \& Herbst 1994, Millar et al. 2000) predict significant abundance
for cyanoplyynes only in the outer part of the envelope, i.e a hollow-shell spatial distribution. That prediction 
is broadly consistent with our observations.

Interestingly, and for the first time, we can trace a series of arcs in the channel maps
of both HC$_3$N and HC$_5$N around the systemic velocity of the envelope, i.e spatially located close to
the plane of the sky. 
In Figure 3, we identify and sketch the location of seven such arcs traced by both cyanopolyyne molecules. 
The location of these arcs can be seen to coincide very closely with the dust arcs 
identified by Mauron \& Huggins
(1999, 2000) and more recently by Le\~{a}o et al. (2006). 
Our observations therefore show that the arcs are physical
structures in the inner envelope and not the projection on the plane of the sky of the larger molecular shells 
seen by Fong et al. (2003). The good spatial correspondance between molecular and dust arcs also 
indicates that the molecular shells observed here
are not related to any peculiarities in the chemistry of the cyanopolyyne molecules but represent true density
variations in the envelope of IRC+10216.

As can be seen in the channel maps, some of the molecular arcs clearly
do not span the whole velocity range of the envelope, e.g arcs  II and VII. Therefore, the molecular arcs represent
incomplete expanding shells in the envelope of IRC+10216. We also note the shapes of
the molecular shells identified in our observations are not exactly circular, (e.g most prominently shell VI in Figure. 3), 
as expected for spherically symmetric and expanding shells,
suggesting that there might be slight variations in ejection velocity or in ejection time between different 
parts of the shell. Furthermore, the spacings between
shells are also not regular, ranging from about $\sim$3 arcsec up to $\sim$9 arcsec. At an expansion velocity of 14.5 kms$^{-1}$ 
(Lucas \& Gu\'{e}lin 1999) and distance of 120 pc, 
the kinematic timescale between the shells is in the range $\sim$120 yrs to $\sim$360 yrs. 

In Figure 4, we show the position-velocity diagram of HC$_3$N J=5$-$4 emission nearly along the North-South direction (PA=$-$25$^\circ$)
and along the direction at position angle PA=45$^\circ$. 
Near the systemic velocity of $-$26 kms$^{-1}$ there are clearly 4 shells visible in the cut at position angle PA=$-$25$^\circ$
and 5 shells in the cut at position angle PA=45$^\circ$. For comparison
purposes, we also show the expected size as a function of radial velocity, 
$r(V)=R_{\rm shell}[1\,-\,(V\,-\,V_{*})^2/V_{\rm exp}^2]^{1/2}$, of a shell with true radius R$_{\rm shell}$, an expansion velocity
$V_{\rm exp}$ of 14.6 kms$^{-1}$, and a systemic velocity V$_*$. By matching the observed size to the prediction we find a radius
of 25 arcsecs for shell VII and a radius of
16 arcsecs for shell VI, respectively (see Fig. 4). We estimate an angular size of $\sim$60$^\circ$ for shell VII in the line of sight direction,
based on the velocity range spanned by this shell from $-$32 kms$^{-1}$ to $-$18 kms$^{-1}$. The angular size of the shell VII 
in the plane of the sky, as can be seen directly in the channel maps at the systemic velocity, is also about $\sim$60$^\circ$. Thus, 
the solid angle subtended by the shell VII is $\sim$1 steradian. Other shells are also found to have comparable size.
As a result, provided their lateral expansion has been negligible, the shells cover roughly about 
10\% of the stellar surface at the time of their ejection.

The good spatial correspondance between HC$_3$N and HC$_5$N, as shown in our VLA maps (see Fig. 5), has been noted earlier by 
Lucas \& Gu\'{e}lin (1999) but at higher frequency transitions in the 3mm band. The latter is less radially extended and traces 
only the inner part of the overall hollow quasi-spherical structure, i.e shells II, IV and the northern end of VI. 
The higher frequency transitions are therefore excited only in the inner and denser part of the envelope and do not trace the full
distribution of cyanopolyyne molecules.
The shells shown in Figure 3 seem to be more concentrated on the western part of the envelope and scarce
at position angles of $\sim$ 0$^\circ$ to 30$^\circ$
(measured counterclockwise from north). Previous observations by Gu\'{e}lin et al. (1993) and
Lucas \& Gu\'{e}lin (1999) also show a lack of molecular emission, 
interpreted as a gas density minimum in that part of the envelope.  
Apart from these asymmetries, the shells are distributed more or less randomly around the
central star. This asymmetry results in a displacement between the 
centroid of the overal hollow shell-like structure 
and the position of the central star as previously noted in
lower angular-resolution observations (Gu\'{e}lin et al. 1993).

\section{Discussion and Conclusion}
Our observations of the molecular envelope of IRC+10216 has the highest angular resolution so far achieved,
and clearly establish 
the correspondance between dust arcs seen previously in scattered light and the molecular shells traced by
the emission of cyanopolyyne molecules.
These are true 3-dimension expanding features located in the inner envelope, and not projections
on the plane of the sky of larger shells as suggested by Fong et al. (2003). 
With kinematic information, we also show that 
the molecular (and by implication the dusty) shells seem to occur randomly in space. Our data also provide
the lower limit to the gas density variation within the envelope. If we assume that
abundance of cyanopolyynes is roughly constant within the emitting region as suggested in the modelling
work of Brown \& Millar (2003), the density contrast between the gas in the molecular shells and the
surrounding gas would correspond to the variation of the intensity of cyanopolyyne emission. A quick 
inspection of Figure 3 shows that the density contrast is at least a factor of a few up to 10 for the
part of the envelope with strongest cyanopolyyne emission.
We note that Mauron \& Huggins (1998) also inferred that 
the gas density in the dusty shells is enhanced by up to an order of magnitude with respect to the intershell medium,
similar to our rough estimate here.
These shells therefore represent brief episodes of enhanced mass loss from the central star separated by 
irregular intervals of a few hundreds years. The similar radial distance of some shells, such as
II and VII, suggest, however, that they are ejected at more or less the same time and, as a result, probably 
belong to the same episode of mass loss enhancement. The presence of molecular shells with significant 
density enhancement within the envelope of IRC+10216 is a clear evidence that the mass loss from the central
AGB star is neither isotropic nor steady over time.

At the present time, it is still difficult to identify any mechanisms that can create the observed
non-isotropic mass loss enhancement over intervals of time much longer than normal stellar pulsation of
649 days for IRC+10216 (Le Bertre 1992), and at the same time much shorter than the interval between helium flashes. By analogy
with solar cycle, Soker (2000) suggests that the magnetic cycle of AGB stars might be responsible for the
observed phenomenon. According to Soker (2000) during the active phase, magnetic cool spots (stellar spots)
appear and reduce the gas temperature of the stellar atmosphere above the spots. That could enhance dust formation
and thus lead to higher mass loss from magnetic cool spots. The sporadic appearance and random distribution of
magnetic cool spots on the stellar surface might also explain the clumpiness seen in the dust arcs and their obviously
random spatial distribution. 

Our VLA observations (see Figure 5) show that the spatial distribution of cyanopolyyne molecules is
very similar, even at an angular resolution of 1.5 arcsec, corresponding to a linear scale of 200 AU. 
The good spatial correlation between the molecular shells and the dusty arcs seen in scattered light
strongly suggests a close coupling between cyanopolyyne-related chemistry and the density enhancements
represented by these shells. The more recent chemical model of Brown \& Millar (2003) suggests that by 
including explicitly the density enhancements in 
the envelope of IRC+10216 the chemically active region tends to narrow and the model
reproduces better the observed close spatial correspondance of cyanopolyyne molecules. Thus, our observations 
of HC$_3$N and HC$_5$N are in qualitative agreement with the model of Brown \& Millar (2003).

We note the presence of strong emissions of HC$_3$N and HC$_5$N
in the outer shells, such as shell I and VII, where previously only emission from the CN radical 
is seen (Lucas \& Gu\'{e}lin 1999).
The CN radical, which is the photodissociation product of the parent molecule HCN, is predicted by the chemical 
models of Cherchneff et al. (1993) and Millar \& Herbst (1994) to
have the most radially extended spatial distribution, whereas the distribution of cyanopolyyne molecules is
predicted to be more sharply peaked with radius and spatially much less extended than that for CN radical.
Therefore, chemical models previously tailored for IRC+10216 can not account for the similarity in
spatial distribution of cyanopolyyne molecules and CN radical.
Clearly more sophisticated chemical models, which include explicitly the sporadic mass loss enhancement and
complex envelope structure, should be explored to better understand the formation of cyanopolyyne molecules
and their relation to other chemically important and very abundant molecules such as CN.

We thank the referee, Dr. A.J. Remijan, for constructive comments, which help to improve our paper.
We are also grateful to Dr. I.C. Le\~{a}o for providing the V-band image of IRC+10216.
This research has made use of NASA's Astrophysics Data System Bibliographic Services
and the SIMBAD database, operated at CDS, Strasbourg, France.

\newpage

\begin{table}
\caption{Summary of the VLA observations}
\begin{tabular}{lcclc}\hline
Line            &   Frequency  &  Obs. mode       & Synthesized beam   & rms (mJy/beam) \\ 
                &     (GHz)    &                  &                    &  ($\Delta$V = 3 kms$^{-1}$) \\ \hline
HC$_3$N J=$5-$4   &   45.490316        &  7-field mosaic  &  2.0"x1.6" PA=5$^{\circ}$       &   3.6         \\
HC$_5$N J=$16-$15 &   42.602171        &  7-field mosaic  &  2.4"x1.7" PA=$-$48.6$^{\circ}$ &   1.4         \\
HC$_5$N J=$9-$8   &   23.963897        &  single field    &  3.4"x3.1" PA=8.6$^{\circ}$     &   1.0        \\ \hline
\end{tabular}
\end{table}

\newpage

\begin{figure*}[ht]
\plotone{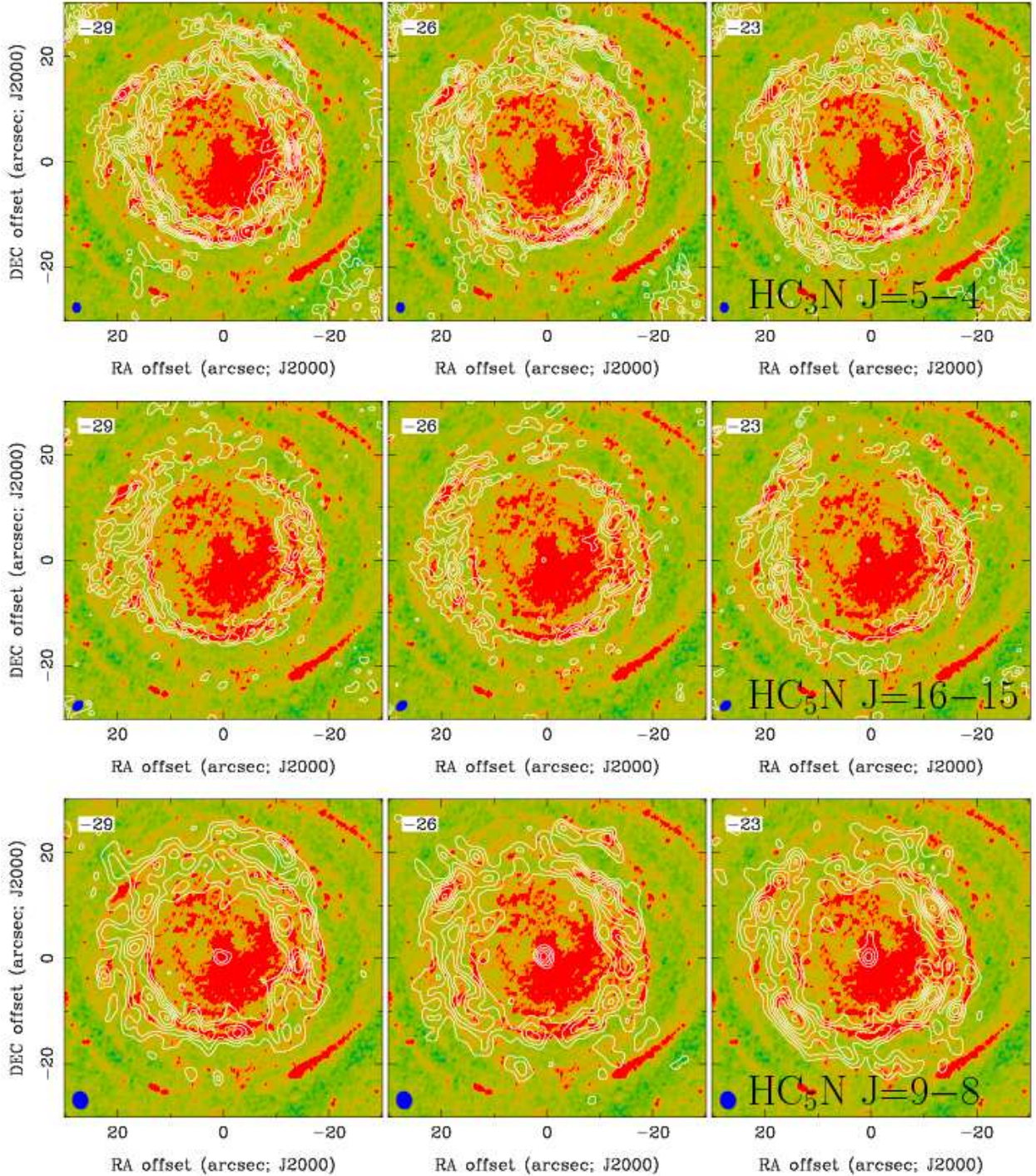}
\caption{Channel maps of cyanopolyyne emission in contours superposed on optical V-band image (Le\~{a}o et al. 2006) of IRC+10216 in false color. 
HC$_3$N J=$5-$4 emission is shown in the upper frame,
HC$_5$N J=16$-$15 in the middle frame and HC$_5$N J=9$-$8 in the lower frame, respectively. Channel velocities are indicated
in the upper left corner. The synthesized beams are shown
in the lower left corner. Contour levels are (3, 5, 7, 9, 11, 13)$\sigma$ for 
HC$_3$N J=$5-$4 ($\sigma$=3.6 mJy/beam) and HC$_5$N J=$16-$15 ($\sigma$=1.4 mJy/beam),
respectively. Contour levels are (2, 3, 4, 5, 6, 7, 9, 11)$\sigma$ for HC$_5$N J=$9-$8 ($\sigma$=1.0 mJy/beam).}
\label{fig1}
\end{figure*}

\newpage

\begin{figure*}[ht]
\plotone{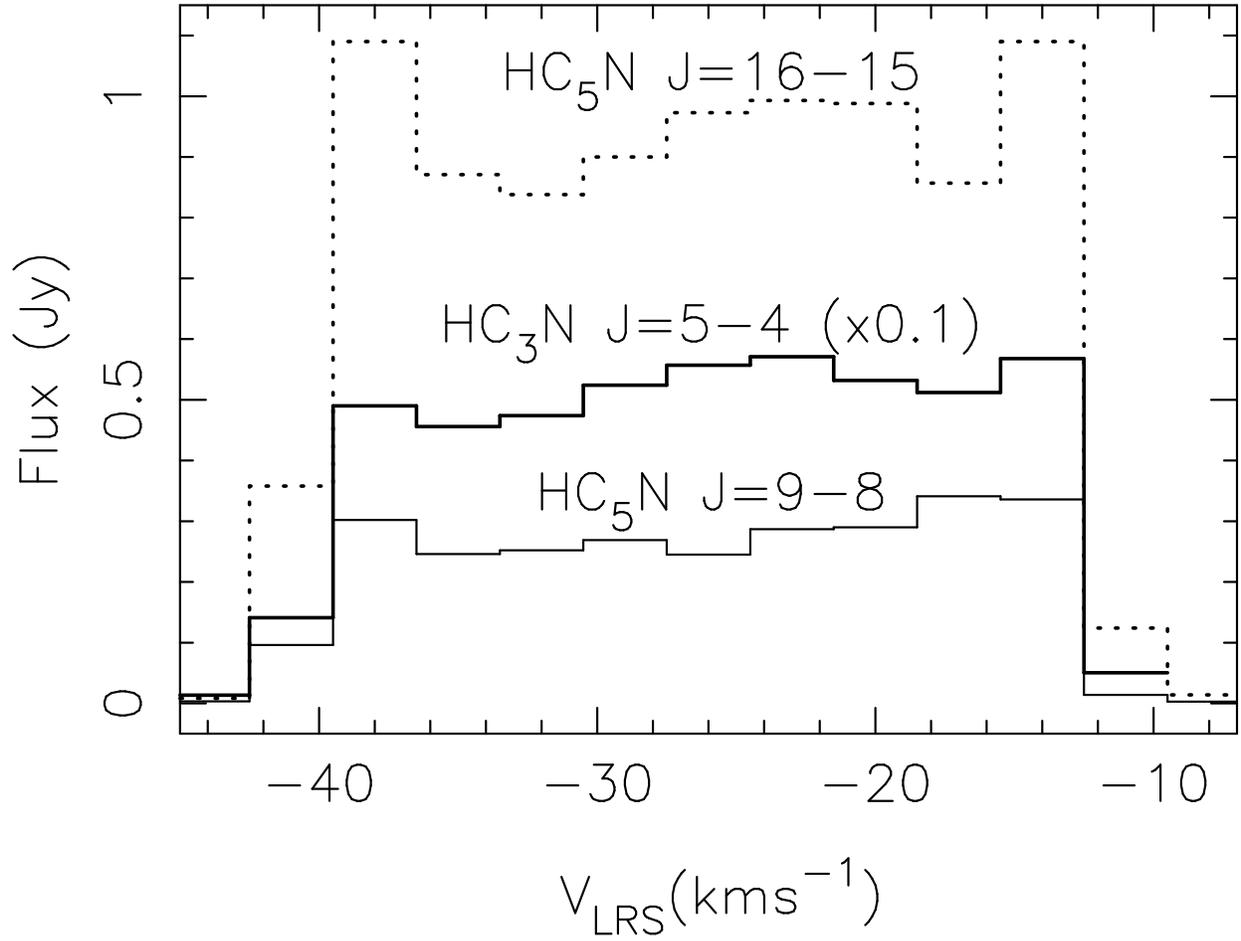}
\caption{The total flux profiles of HC$_3$N J=$5-$4 (scaled by a factor of 0.1),
HC$_5$N J=$9-$8 and HC$_5$N J=$16-$15 lines.}
\label{fig2}
\end{figure*}

\newpage

\begin{figure*}[ht]
\plotone{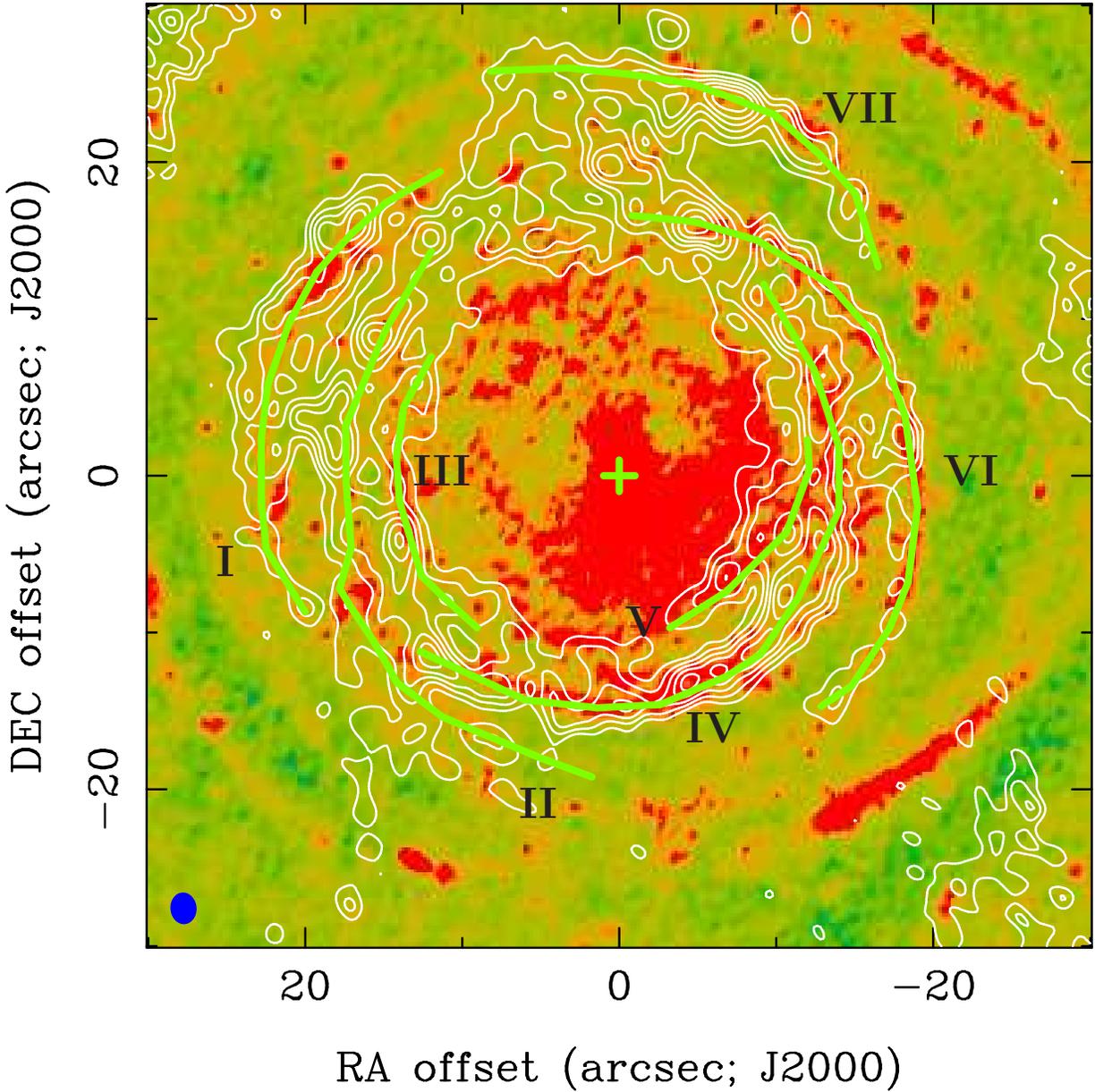}
\caption{Location of the molecular shells (thick solid lines) traced by cyanopolyyne emissions around the systemic velocity, i.e close
to the plane of the sky. The cross denotes the central stellar position.
The shells are numbered I - VII. The HC$_3$N J=$5-$4 emission at the systemic velocity is shown in contours.
Contour levels are (3, 5, 7, 9, 11, 13)$\sigma$ with $\sigma$=3.6 mJy/beam}
\label{fig3}
\end{figure*}

\newpage

\begin{figure*}[ht]
\setlength{\unitlength}{1cm}
\begin{picture}(10.0,17.5)
\put(0,0.){\resizebox{14.cm}{!}{\includegraphics*{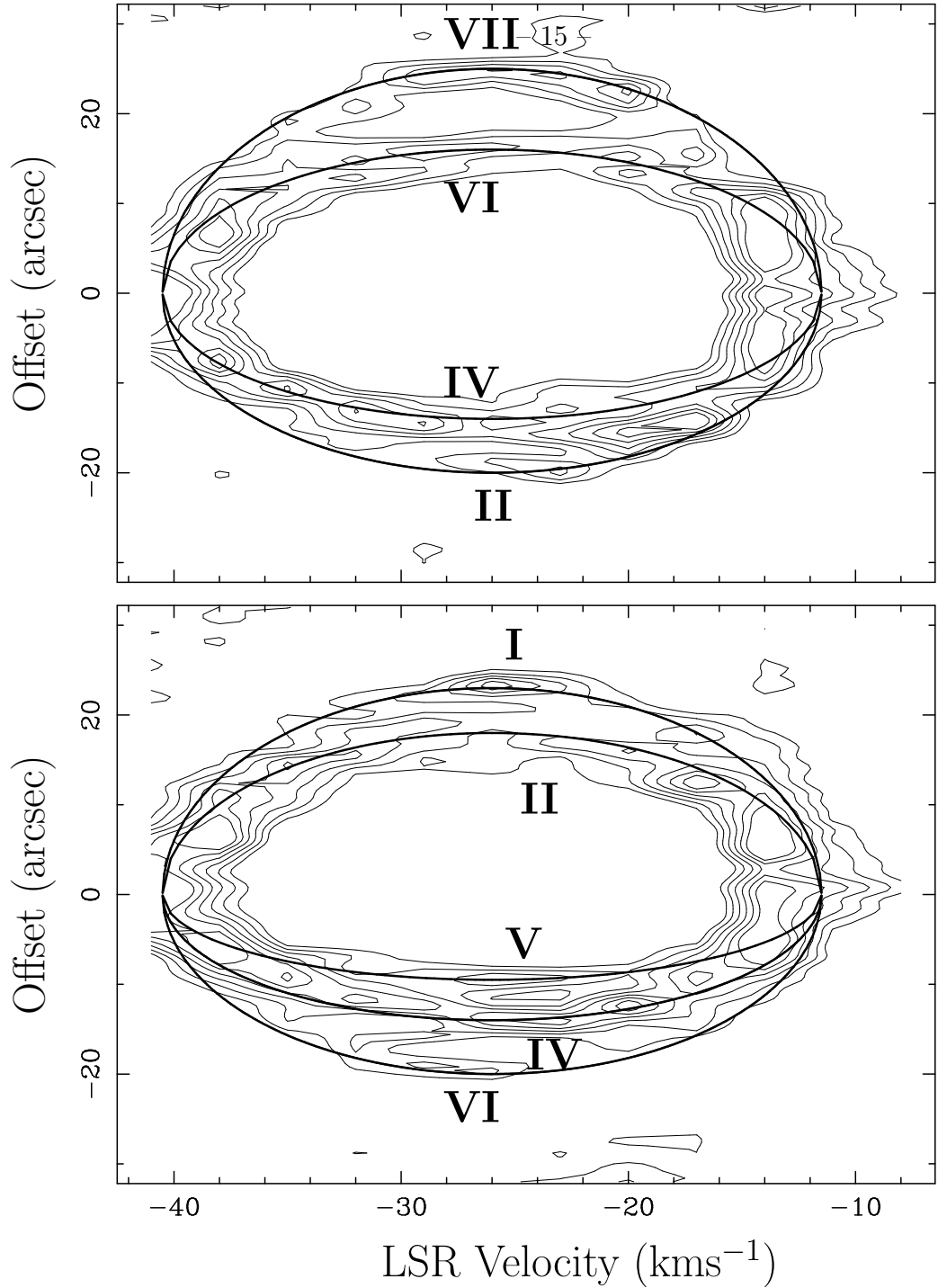}}}
\end{picture}
\caption{{\bf Upper frame:} position-velocity map of HC$_3$N J=$5-$4 along a cut at position angle PA=$-$25$^\circ$ through 
shells numbered VII, VI, IV and II. Thick solid lines represent (from the top to bottom)
the 04 shells with radius 25 arcsec, 16 arcsec, 14 arcsec and 20 arcsec, respectively. 
{\bf Lower frame:} position-velocity map of HC$_3$N J=$5-$4 along a cut at position angle PA=45$^\circ$ through arcs 
numbered I, II, V, IV, and VI. 
Contour levels are (3, 5, 7, 9, 11, 13)$\sigma$. Solid lines represent (from the top to bottom)
the five shells with radius 23 arcsec, 18 arcsec, 9.5 arcsec, 14 arcsec and 20 arcsec, respectively.
In both frames, contour levels are (3, 5, 7, 9, 11, 13)$\sigma$ with $\sigma$=1.0 mJy/beam.
The expansion velocity is assumed to be 14.5 kms$^{-1}$}
\label{fig4}
\end{figure*}

\newpage

\begin{figure*}[ht]
\plotone{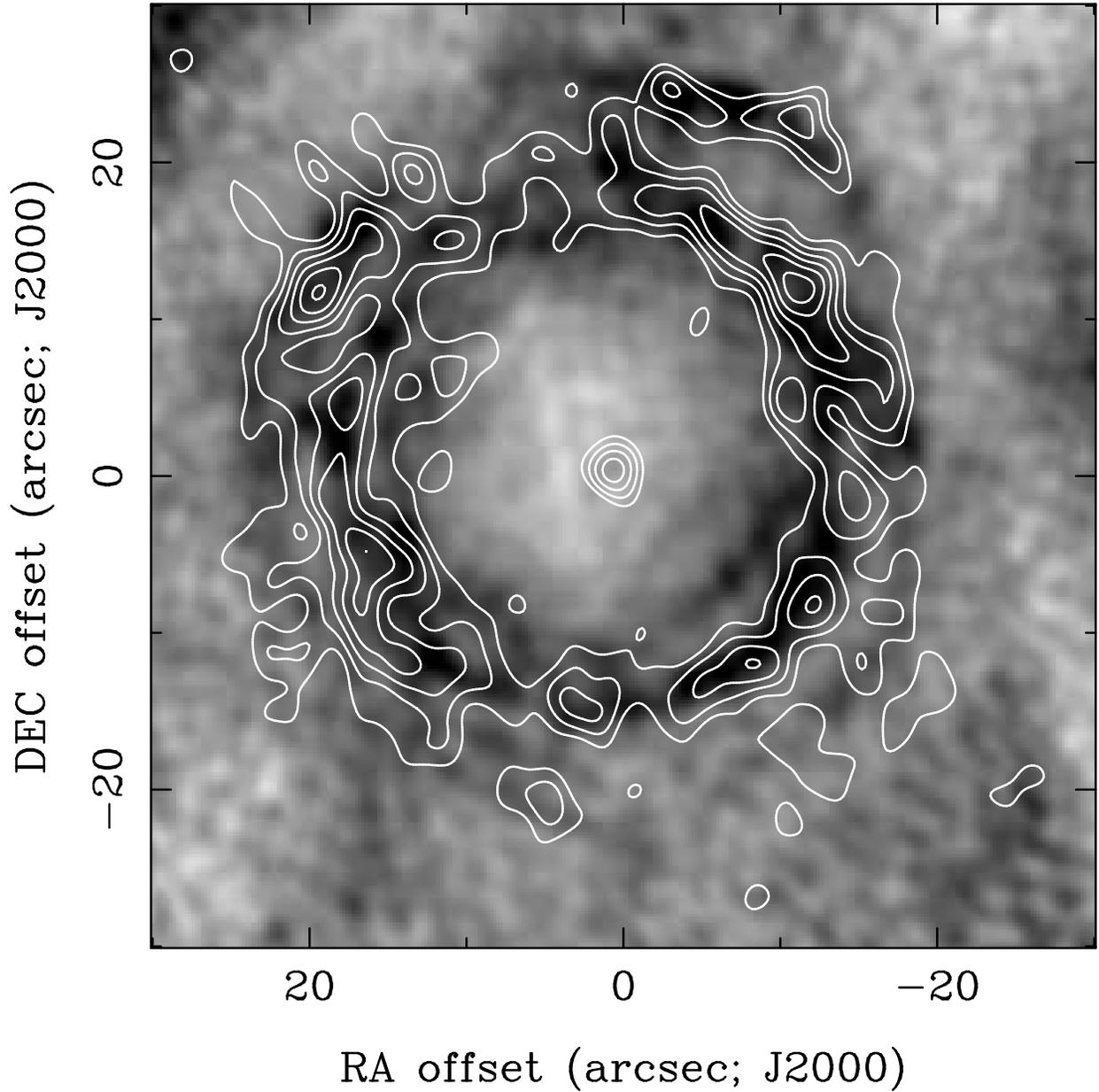}
\caption{Comparison between HC$_3$N J=$5-$4 emission (shown in greyscale) and HC$_5$N J=$9-$8 emission (shown in contours) at
the systemic velocity V$_{\rm LSR}$ = $-$26 kms$^{-1}$. The contour levels are 
(2, 3, 4, 5, 6, 7, 9, 11)$\sigma$ for HC$_5$N J=$9-$8 ($\sigma$=1.0 mJy/beam).}
\label{fig5}
\end{figure*}

\end{document}